\documentclass[reprint,amsmath,aps,pra,footinbib,notitlepage,longbibliography,superscriptaddress]{revtex4-2}
\usepackage{graphicx,epstopdf,dcolumn,bm,mathtools,siunitx,multirow,hyperref,xcolor,amsfonts,physics,bibunits}

\begin{document}
\title{Quantum acoustics unravels Planckian resistivity}
\thanks{Notice: This is an updated version of the manuscript after acceptance. The final published version can be found in the Proceedings of the National Academy of Sciences (PNAS) website at the following link: \href{https://www.pnas.org/doi/10.1073/pnas.2404853121}{https://www.pnas.org/doi/10.1073/pnas.2404853121}}
\author{Alhun Aydin}
\affiliation{Department of Physics, Harvard University, Cambridge, Massachusetts 02138, USA}
\affiliation{Faculty of Engineering and Natural Sciences, Sabanci University, 34956 Tuzla, Istanbul, Turkey}

\author{Joonas Keski-Rahkonen}
\affiliation{Computational Physics Laboratory, Tampere University, P.O. Box 692, FI-33101 Tampere, Finland}
\affiliation{Department of Physics, Harvard University, Cambridge, Massachusetts 02138, USA}
\affiliation{Department of Chemistry and Chemical Biology, Harvard University, Cambridge,
Massachusetts 02138, USA}

\author{Eric J. Heller}
\affiliation{Department of Physics, Harvard University, Cambridge, Massachusetts 02138, USA}
\affiliation{Department of Chemistry and Chemical Biology, Harvard University, Cambridge,
Massachusetts 02138, USA}

\date{\today}
\begin{abstract}

Strange metals exhibit universal linear-in-temperature resistivity described by a Planckian scattering rate, the origin of which remains elusive. By employing a novel approach inspired by quantum optics, we arrive at the coherent state representation of lattice vibrations: quantum acoustics. Utilizing this nonperturbative framework, we demonstrate that lattice vibrations could serve as active drivers in the Planckian resistivity phenomenon, challenging prevailing theories. By treating charge carriers as quantum wave packets negotiating the dynamic acoustic field, we find that a competition ensues between localization and delocalization giving rise to the previously conjectured universal quantum bound of diffusion, $\hbar/m^{*}$, independent of temperature or any other material parameters. This leads to the enigmatic $T$-linear resistivity over hundreds of degrees, except at very low temperatures. Quantum diffusion also explains why strange metals have much higher electrical resistivity than typical metals. Our work elucidates the critical role of phonons in Planckian resistivity from a new perspective and reconsiders their significance in the transport properties of strange metals.

\end{abstract}
\maketitle



\section{Introduction}\label{s:intro}

Understanding the mechanism behind high-temperature (high T$_c$) superconductivity remains one of the central questions in condensed matter physics, ever since its discovery almost four decades ago~\cite{Keimer2015}. The high-T$_c$ materials —such as cuprates, heavy-fermion compounds, iron-based pnictides, and twisted bilayer graphene— exhibit anomalous resistivity behaviors above their superconducting transition temperature, earning them the name of "strange metals". It is thought that finding the mechanisms behind this unusual resistivity will provide key insights into their superconductivity~\cite{Varma2020}. Universal linear temperature ($T$-linear) dependence of resistivity over a wide temperature range~\cite{Varma2020,hussey11,spott,strangemetal1,Legros2018,Polshyn2019,Greene2020,Grissonnanche2021,PhysRevB.106.035107}, violation of Mott-Ioffe-Regel (MIR) limit~\cite{tahir-kheli_resistance_2013} and the build-up of displaced Drude peaks (DDPs) in the optical conductivity~\cite{infrared,infrared2} are among major still unexplained transport properties of strange metals.

Fermi liquid theory suggests $T^2$ dependence for the resistivity of metals due to electron-electron interactions~\cite{Many-Particle,behnia}. Bloch-Gr{\"u}neisen theory, on the other hand, predicts $T^{2+d}$ ($d$ is the dimension of the system) at low temperatures and $T$-dependence at high temperatures from electron-phonon scattering~\cite{ziman2001electrons}. Contrary to these conventional theories, the resistivity of strange metals is orders of magnitude larger than that of typical metals and exhibits a $T$-linear dependence starting from the superconducting transition temperature up to the melting temperatures~\cite{Varma2020}. Even more mysteriously, the slope of the $T$-linear resistivity obeys the so-called Planckian scattering rate
\begin{align}
\tau_P=\alpha\frac{\hbar}{k_BT},
\end{align}
where $\hbar$ is Planck constant, $k_B$ is Boltzmann constant and $\alpha$ is Planckian coefficient, a material-dependent constant in the order of unity, which appears to be universally applicable to a wide range of quantum materials~\cite{strangemetal1,Legros2018,Zaanen2019,Grissonnanche2021,planckianmetals}. Calculating resistivity via the Drude formula, $\rho=m^*/(ne^2\tau_P)$  using the Planckian scattering rate for $\tau_P$ perfectly predicts the experimental data~\cite{strangemetal1,Legros2018,planckianmetals}. This suggests that a single scattering mechanism governs the transport behavior in these materials, independent of their internal structure~\cite{strangemetal1,Legros2018}. The fact that Planckian properties have been found in 2D semiconductors also implies that the underlying cause of the Planckian resistivity could indeed be captured by a universal mechanism~\cite{PhysRevB.106.155427}. The Planckian scattering rate is also conjectured to be a universal bound on scattering rates in thermal systems, regardless of the underlying mechanism~\cite{Zaanen2019,planckianmetals}. The bound is motivated and obtained in an {\it ad hoc} manner by invoking time-energy uncertainty relation, $\Delta t\Delta E\approx \hbar$, supposing the characteristic energy scale to be $k_BT$. It appears in a variety of different contexts in physics, from black holes to quantum chaos~\cite{Zaanen2019,planckianmetals}, adding to the  mystery of this ubiquitous phenomenon. 

Many-body electron-electron interaction has been thought to be the main cause for   strange metal behavior since it can give rise to emergent phenomena missed by  conventional single-particle theories~\cite{Sachdev,RevModPhys.94.035004}. While angle-resolved photoemission spectroscopy (ARPES) experiments show the ubiquitous strong electron-phonon coupling in cuprates~\cite{Lanzara2001}, the resistivity predictions of conventional electron-phonon interactions don't match the experimental data, particularly at low temperatures. This and other unconventional superconductivity-related discrepancies caused many to conclude that phonons could not play the leading role in the resistivity of strange metals. On the other hand, the significant influence of electron-phonon interactions in cuprates has been evidenced by the pronounced isotope effect~\cite{PhysRevLett.86.4899} and the T-linear resistivity observed in NbSe3 above its upper Peierls transition~\cite{Yang2019}. Recently it has been argued that phonons alone could predict $T$-linear resistivity in strange metals as consequences of lattice inhomogeneity~\cite{tahir-kheli_resistance_2013}, low charge carrier density~\cite{hwangSarma2019}, collective modes~\cite{PhysRevB.88.115103}, flat bands, and suppression of Fermi velocity~\cite{Lanzara2001,PhysRevB.99.165112,Polshyn2019,PhysRevB.106.035107,PhysRevB.107.235155}. Vortex states are found in cuprates, suggesting that the conventional phonon-assisted pairing mechanism for superconductivity could be at play in strange metals as well~\cite{PhysRevLett.119.237001}. Although there were several partially successful approaches~\cite{doi:10.1073/pnas.2003179117,smsubir}, none of them provide a complete resolution for the strange metal mysteries within a single unified picture.

Here we propose and verify numerically that Planckian resistivity can be manifested exclusively through interactions between charge carriers and lattice vibrations under a previously unexplored, coherent and emergent quantum diffusive regime. Using a real-space and time-dependent Wave-on-Wave (WoW) formalism, and adopting the coherent state representation for lattice vibrations (quantum acoustics), we unveil coherence-preserving and nonperturbative scattering dynamics. Our transport simulations reveal a diffusion of carriers constrained by a previously conjectured quantum bound~\cite{Hartnoll2015,Hu2017,PhysRevLett.119.141601}. We validate this model by accurately reproducing both the magnitude and the slope of resistivity for three distinct strange metals using their empirical parameters near the critical doping and above the critical temperature. Our results predict T-linear resistivity with the Planckian slope starting from around superconducting transition temperatures up to melting temperatures. Our findings remain fairly robust against reasonable variations in material parameters. Additionally, we offer a phenomenological framework elucidating the quantum diffusion-based mechanism responsible for Planckian scattering. Our work could shed light on other persisting strange metal mysteries. In particular, based on our model, we show that Mott-Ioffe-Regel (MIR) limit is surpassed and a shift in the Drude peak appears in the optical conductivity at high temperatures.

\begin{figure*}[t]
\centering
\includegraphics[width=0.8\textwidth]{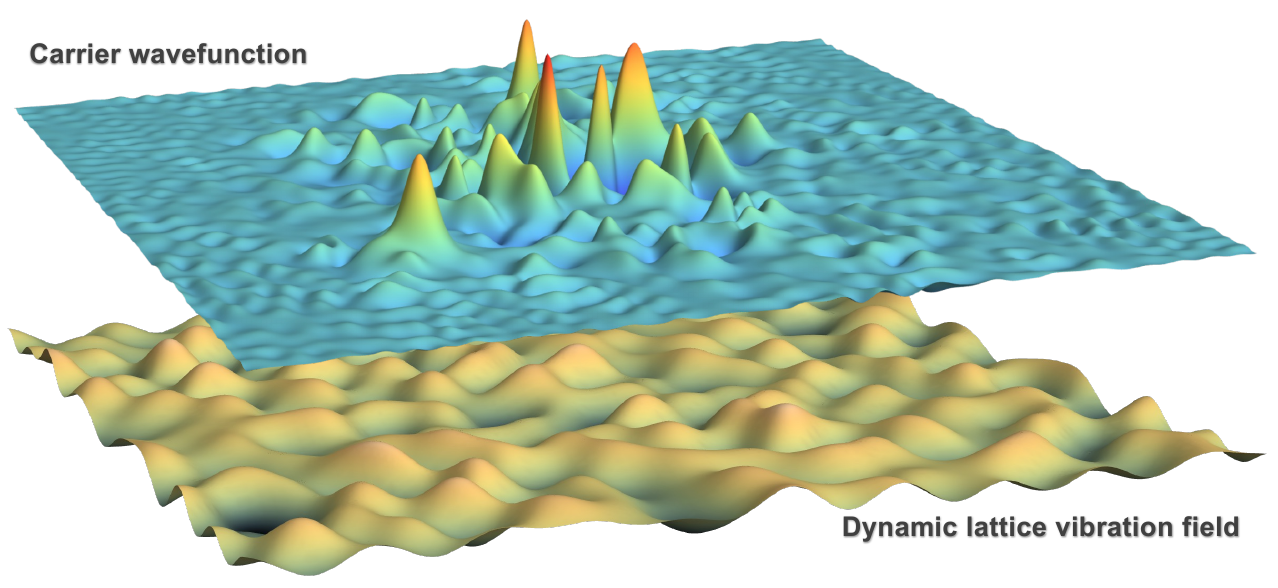}
\caption{Wave-on-Wave approach. Nonperturbative and coherent dynamics of charge carriers and thermal lattice vibrations. A snapshot of a carrier wave packet (the real part is shown on top) coherently propagating under a spatially continuous, dynamic disorder field (bottom) formed by acoustic lattice deformations. A charge carrier quasielastically scatters (similar to impurity scattering) in this disordered landscape. There is no single characteristic wavelength in both fields but they possess average length scales making it possible to distinguish the coherent/incoherent behaviors.}
\label{coherent}
\end{figure*}

\section*{Nonperturbative and coherent carrier scattering in strange metals}

Temperature-dependent resistivity is mainly determined by the scattering of charge carriers with longitudinal acoustic phonons~\cite{ziman2001electrons}. To describe phonons, we use a recently developed~\cite{Heller22} coherent state representation and refer to disturbances as \textit{lattice vibrations}, as opposed to individual \textit{phonons}, emphasizing their coherent wave nature, in exact parallel to the electromagnetic field limit of quantum optics.  The powerful coherent state representation enables us to construct a quasiclassical field of lattice vibrations, \emph{deformation potential} in coordinate space, allowing a quantum-coherent treatment of charge carriers as well as nonperturbative scatterings. The spatial coherence of carriers is preserved beyond single collision events~\cite{Heller22} which is one of the key new features of our approach (see Materials and Methods \ref{s:Wow} for further details).

In the usual Boltzmann transport treatment, starting with perturbation theory, carrier transport is considered incoherent and perturbative, which works well for typical metals~\cite{ziman2001electrons}. In strange metals, on the other hand, the quasiparticle picture breaks down and carriers cannot be considered as semiclassical entities~\cite{stranger} subject to Boltzmann diffusion. Charge carriers acquire longer wavelengths compared to the characteristic length scale of the deformation potential, putting them into the quantum-coherent regime where quantum interference effects become important~\cite{dattabook1995}. Strong deformation potential coupling and small Fermi energy of most strange metals put them into the nonperturbative scattering regime where the electron-phonon coupling cannot be considered as a perturbative correction to the electronic band energy~\cite{Lanzara2001}. This indicates that the role of phonons may extend to lower temperature ranges than traditionally thought~\cite{Mousatov2021}. 

We associate the nonperturbative and coherent dynamics with the energy and wavelength of carriers being larger or at least comparable to the strength of the deformation potential and the lattice constant. The strange metals under consideration belong to this category due to their low Fermi energy and strong electron-lattice coupling, see Materials and Methods. We carefully treat the quantum nature of the carrier scattering by considering the quantum interference effects in coordinate space~\cite{dattabook1995}. We describe the charge carriers as quantum wave packets moving under the dynamic lattice wave field arising from the standard Fr{\"o}hlich Hamiltonian~\cite{Frochlich1, Frochlich2, Mahan}, which we call the WoW method, explained further in Materials and Methods \ref{s:Wow}. We illustrate the snapshot of the real part of the carrier wave function (top) evolved after some time under the deformation potential (bottom) in Fig. \ref{coherent}. Carrier wave functions undergo phase-preserving multiple quasielastic scatterings everywhere at once until they lose their coherence in real space. Scatterings are ubiquitous, frequent, and local (short-range). Note that the coherence in energy basis could have been lost (thermalization) much earlier~\cite{planckianmetals}. A quantum treatment is necessary for modeling the transport of charge carriers in strange metals, even at high temperatures~\cite{Heller22}. Note that the importance of quantum coherence of carriers at high temperatures has also been highlighted in the context of charge density waves~\cite{PhysRevB.87.115127,10.1063/5.0048834,MILLER2024101326}.

\section*{Quantum-acoustical origin of the Planckian resistivity}

Using the WoW framework, we conduct numerical transport simulations for three different cuprates, LSCO, Bi2212, and YBCO, by considering their material parameters given in Materials and Methods section \ref{s:Parameters}. Material parameters of strange metals suggest that the deformation potential scattering is nonperturbative and charge carriers are coherent throughout the transport time scales. Transport of charge carriers under this coherent, non-perturbative regime is diffusive. We evaluate the instantaneous diffusivity
\begin{align}
D(t)=\frac{1}{2d}\frac{\partial \text{MSD}_\mathbf{x}(t)}{\partial t},
\end{align}
where $d$ is the dimension of the system which is set to two because the carrier transport in cuprates is effectively two-dimensional, taking place in $\text{CuO}_2$ planes. $\text{MSD}_\mathbf{x}(t)$ is the quantum-mechanical spread (mean square distance) of the wave packet in the transport (+x) direction,
\begin{equation}
\text{MSD}_\mathbf{x}(t) =  \int \text{Re} \{ \psi^*(\mathbf{r},t) [\mathbf{x}(t) - \bar{\mathbf{x}}(t)]^2 \psi(\mathbf{r},t) \} \, dx \, dy,
\end{equation}
where $\psi(\mathbf{r},t)$ is the wave function in real space at time $t$, $\mathbf{r}$ is the 2D position operator, $\mathbf{x}$ is the position operator in the transport direction and $\bar{\mathbf{x}}$ is the average distance travelled in the transport direction. The usual diffusion coefficient is the long time limit of instantaneous diffusivity, $D=\lim_{t \to \infty}D(t)$.
In line with the widely acclaimed arguments for strange metals~\cite{strangemetal1,Legros2018,Grissonnanche2021}, we assume a single dominant scattering rate and define carrier mobility phenomenologically as $\mu_c=e\tau/m^{*}$. Drude mobility enters into the Einstein relation for diffusion of charged particles ($D=\mu_ck_BT/e$) giving the inverse scattering rate in terms of diffusion coefficient as
\begin{align}
\frac{1}{\tau}=\frac{k_BT}{m^{*}D}
\end{align}
where $m^{*}$ is the effective mass. This relationship appears to be broadly applicable not only in non-degenerate conductors but also in degenerate systems~\cite{datta_book}.

\begin{figure*}[t]
\centering
\includegraphics[width=0.95\textwidth]{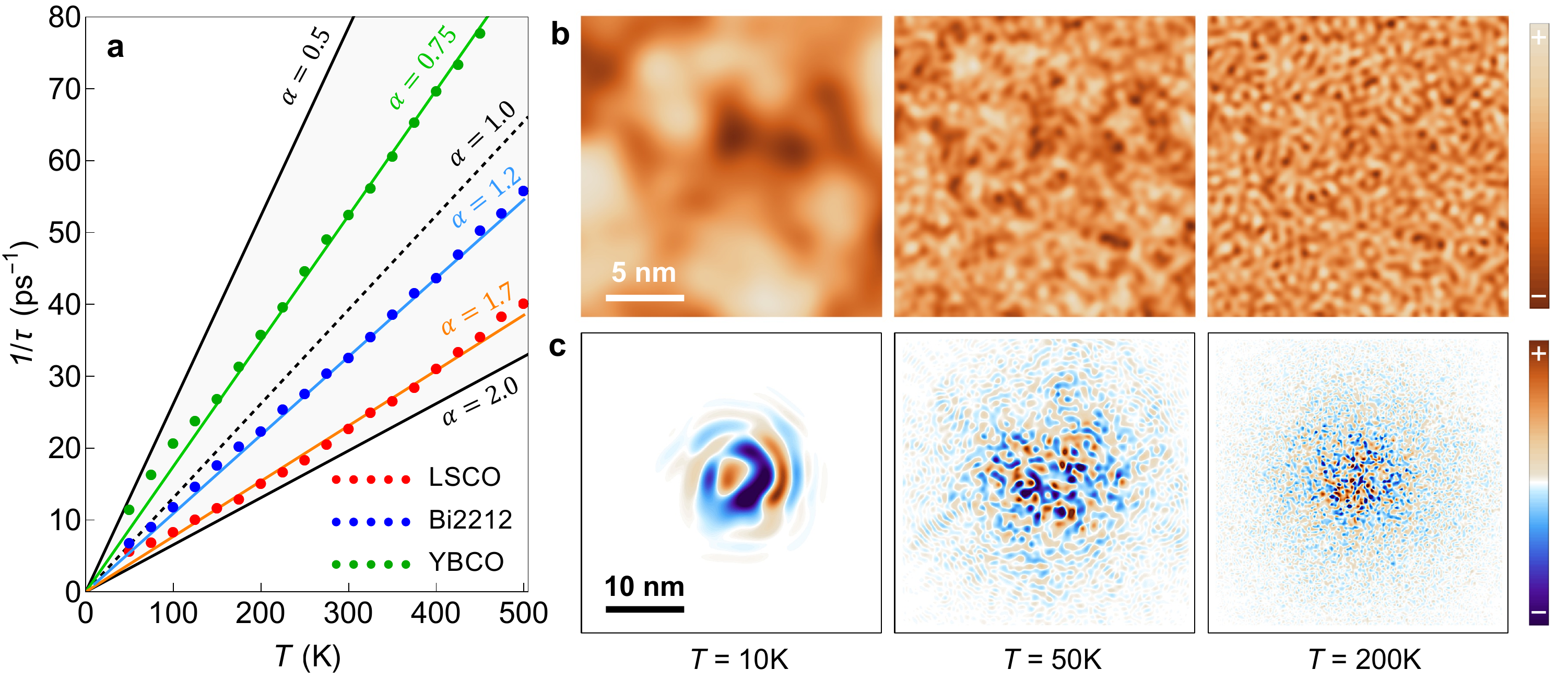}
\caption{Quantum diffusion causes Planckian resistivity. (a) Planckian scattering rates were obtained from charge carrier-lattice vibration dynamics under the emergent quantum diffusion of charge carriers for three different strange metals: LSCO (red), Bi2212 (blue), and YBCO (green). Dots represent our simulation results for the temperature-dependent inverse scattering rates, $1/\tau_P=k_BT/(m^{*}D\alpha)$ with $\alpha$ being a material-dependent empirical parameter. The colored lines show their corresponding Planckian slopes, $1/\tau_P=k_BT/(\hbar\alpha)$. Black lines show the effective $\alpha$ parameter range for strange metals and the dashed line presents the exact Planckian slope, $\alpha=1$. Snapshots of the deformation potentials for LSCO are shown in (b), for three different temperatures, $T=10$K, $T=50$K, and $T=200$K in a row. Real parts of the carrier wavefunctions propagating under these deformation potentials are respectively plotted in (c). Legends are in arbitrary scale, just showing the intensities and signs.}
\label{planckian}
\end{figure*}

By applying the Drude-Einstein phenomenology, we plot the temperature-dependent inverse scattering rates in Fig. \ref{planckian}(a). For the cuprates that are considered, our numerical results and their fitted Planckian slopes are shown with colored dots and lines respectively. Experimental data predicts T-linearity with Planckian coefficient $\alpha$ usually ranging between 0.5 and 2, but mostly near 1, see \ref{s:Parameters}. In other words, the gray region in Fig. \ref{planckian}(a) denotes the range where most strange metals lie. As is shown, we predict the experimental data quite well both in magnitude and in slope starting from as low as 50K up to 500K temperature for these cuprates. Next, we show the snapshots of the underlying deformation potentials in Fig. \ref{planckian}(b), and snapshots of the real parts of carrier wave functions  in Fig. \ref{planckian}(c), for three different temperatures of $T=10$K, $T=50$K and $T=200$K in a row from left to right. At $T=10$K, the deformation potential is not strong enough so that the scattering is perturbative, and coherence effects are almost negligible due to the large deformation potential length scale compared to the carrier wavelength~\cite{Heller22}. Starting with $T=50$K, both the increased magnitude and the emergent, small length scales of the deformation potential make the scattering dynamics nonperturbative and coherent, as is also evident from the snapshots of the carrier wave functions. This T-linearity produced by a nonperturbative and fully quantum model with emergent quantum diffusion is quite remarkable and distinctly different from the T-linear resistivity observed in normal metals, where the deformation potential acts like a static disorder allowing for perturbative scattering treatments~\cite{Heller22}.

Although the Planckian slope depends on many factors such as effective mass, sound velocity, deformation potential constant, and lattice parameters, it predominantly falls within the specified range of the Planckian coefficient. The slope predominantly displays robustness in that sense, however, we also observe multi-slope behavior (each being linear within shorter temperature ranges) at higher deformation potential constants. We observe slight deviations from the expected behaviors at the extreme parameter values, see the ranges we consider in Materials and Methods \ref{s:Parameters}. Note that these parameters could depend on and vary with several other factors, such as doping, band structure, and various couplings. We haven't explicitly accounted for the doping in this work, rather its influence is implicitly accounted for in the considered material parameters. While the slope is also sensitive to the doping value, the order of magnitude of $\alpha$ is the same. Ultimately, the considered material values provide a good overall estimate as our purpose here is to show the robustness of the quantum diffusion within a reasonable range of material parameters.

Several factors are at play in this emergent quantum diffusive behavior. Under the coherent and nonperturbative scattering regime, charge carriers quickly become immersed in the fluctuating field of lattice vibrations, Fig. \ref{coherent}. Our simulations reveal that the carriers acquire the (sound) velocity of the underlying field along with a new, larger carrier wavelength. Note that ARPES data already reveals the broadening of the velocity in cuprates~\cite{PhysRevLett.86.4362} and the existence of thermal vortexes also implies that the carrier velocity is no longer Fermi velocity~\cite{Li_2021}. Hence, acoustic lattice scattering becomes even more enhanced due to this reduced carrier velocity. The large effective mass of charge carriers (usually holes) is also a contributing factor in nonperturbative dynamics, making carriers easier to trap. In the potential dips, quasibound states are formed. Because of the quantum interference effects between many successive collisions, a new transport regime emerges as the interplay of two competing mechanisms from a single source: While the random deformation potential tries to localize carriers, its dynamic nature works against the localization by scrambling the phase information among other things. Under this interplay between localization and delocalization, carriers display a new kind of transport behavior that is associated with quantum diffusion. We observed all these effects in our numerical simulations, see Fig. \ref{coherent}. The diffusion coefficient of carriers in this emergent regime obeys so-called quantum diffusion $D_Q=\alpha\hbar/m^{*}$ with the same material-dependent parameter $\alpha$ of the Planckian scattering rate. Substituting the quantum diffusion into Eq. (4) gives the inverse Planckian scattering rate, showing the phenomenological equivalence of both concepts,
\begin{align}
\left(\parbox{4.2em}{\centering Quantum\\diffusion}\right)\;D_Q=\alpha\frac{\hbar}{m^{*}}\Longleftrightarrow\tau_P=\alpha\frac{\hbar}{k_BT}\;\left(\parbox{4.2em}{\centering Planckian\\scattering}\right)
\end{align}
which we confirm numerically in Fig. \ref{planckian}. The quantum diffusive regime arises from the competition between localization and release due to lattice motion. 
In fact, many strange metals are on the verge of this type of a transient localization associated with strong electron-phonon interaction~\cite{Lanzara2001, Varma2020}. Furthermore, while some claims regarding the temperature dependence of diffusivity have been made previously \cite{hubbardSM}, these results were based on the assumption that the hopping parameter is temperature-independent. However, this assumption may not hold in conditions of strong electron-phonon interactions, which can significantly alter the overlap of wavefunctions.

A quantum bound on diffusivity has already been predicted to play a role in strange metals~\cite{Hartnoll2015,Hu2017,PhysRevLett.119.141601}. Universal behavior due to quantum diffusion in strongly correlated materials has been observed in Fermi gases~\cite{Sommer2011,doi:10.1126/science.1247425,PhysRevLett.118.130405}. Similar bounds have been proposed, based on the local uncertainty relations, in the contexts of universal chaos and speed limits on transport coefficients including the diffusion constant~\cite{NUSSINOV2020114948,NUSSINOV2022168970,zohar23}. Quantum diffusion has been more explicitly reported for wide class of cuprates by scaling the slope of $T$-linear resistivity with London penetration depth~\cite{Hu2017}. The importance of coherence effects and strong diffusive scattering has also  been suspected in cuprates~\cite{Li2018}. There were also studies arguing the clear distinction of the bounds on conventional and unconventional metals~\cite{stabilitybound}. Recently, a quantum bound on diffusivity has been put forward in various systems~\cite{Hartnoll2015} as well as in strange metals~\cite{Hu2017}, importantly as a possible explanation for $T$-linearity by Hartnoll~\cite{PhysRevLett.119.141601}. Equivalence of quantum diffusion and Planckian scattering rate has also been implied for spin diffusivity~\cite{PhysRevLett.118.130405}. Here we confirm the existence of and obedience to this bound due to acoustic phonons, which is revealed under coherent and nonperturbative dynamics. We want to emphasize the observed behavior emerges from the venerable Fr{\"o}hlich Hamiltonian taken to the coherent state representation of quantum acoustics.

At this point, it is important to acknowledge that the Drude-Einstein phenomenology, connecting quantum diffusion to the transport scattering rate, assumes the breakdown of Fermiology in strange metals. We think the strong deformation potential coupling and coherent dynamics could be the primary reasons behind this breakdown. Fermi energies are already quite low in strange metals (leading to at least an order of magnitude lower Fermi velocities compared to typical metals)~\cite{strangemetal1,Mousatov2021}, because they almost universally have smaller carrier concentrations (unless they are overdoped) which decreases their Fermi level~\cite{tahir-kheli_resistance_2013}. Hence, coherent multiple scattering could be decisive in their transport behavior. The formation of quasibound states with quantum diffusive dynamics could be another factor in foregoing the Fermiology. Such states lead to hydrodynamic-like transport, motivating the use of Einstein relation for the diffusion of charged particles ~\cite{Hartnoll2015,Hu2017,PhysRevLett.119.141601}. An argument in favor of using the Einstein relation in this context is also discussed in Ref~\cite{PhysRevLett.107.066605}. In that context, the scattering rate saturates to a value that is independent of energy, due to the rapid equilibration of charge carriers in strange metals. While the Einstein relation could become invalid under the nonequilibrium effects of deep-level charge trapping, eventual recombination releases the trapped carriers in organic semiconductors and recovers the Einstein diffusion. In our model, the breaking of charge localization due to the time-dependent deformation potential plays a similar role, delocalizing the charge carriers, resulting in fast equilibration that restores the diffusive regime. It has also been shown that $T$-linear resistance can be observed in bosonic hosts~\cite{Yang2022}, suggesting that the cause of strange metal behaviors might not necessarily be related to the fermionic nature of the carriers, supporting the idea that degeneracy does not play much of a role in quantum diffusive transport (Fermiology breaks down~\cite{stranger}). Like the isotropic scattering rate~\cite{Grissonnanche2021} and the robustness towards a magnetic field~\cite{magnetic_field,scalemagnetic}, this experimental piece of the puzzle also fits fairly together within our picture of treating lattice vibrations as propagating waves.

To supplement the quantum diffusion analysis, we have also estimated the relaxation time within the Kubo formalism (see the Materials and Methods \ref{s:DDP}). These calculations give the same order of magnitude as the results discussed above. Additionally, our simulations corroborate the experimentally observed isotropy of the Planckian scattering rate~\cite{Grissonnanche2021}, as this independence of directionality actually stems from the properties of the deformation potential, specifically being homogeneously random~\cite{Heller22}.

\begin{figure}[t]
\centering
\includegraphics[width=0.48\textwidth]{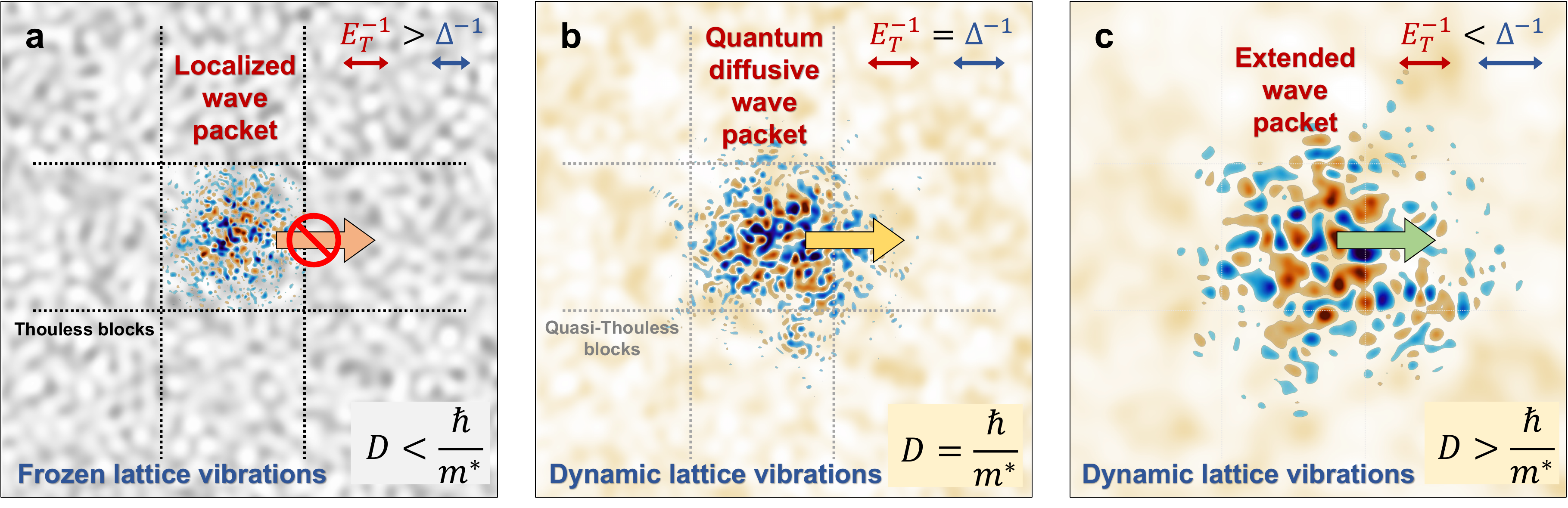}
\caption{Wave packet diffusion in Thouless block (scaling theory) picture. Transition from localized (insulating) to extended (metallic) state is characterized by the comparison of the mean energy level spacing of the wave packet ($\Delta=\hbar^2/m^{*}L^2$) with the Thouless energy ($E_T=\hbar D/L^2$) corresponding to the average energy shift due to diffusion of the wave packet. Equivalently, one can compare the mean lifetime ($\tau_{\Delta}=m^{*}L^2/\hbar$) with average Thouless diffusion time ($\tau_T=L^2/D$) of the wave packet through the (quasi-)Thouless blocks. (a) For a frozen lattice with strong deformation potential, the characteristic energies are $E_T<\Delta$ and the wave packet is localized. (b) At the critical value, $E_T=\Delta$, the diffusion coefficient becomes $D=\hbar/m^{*}$, which is the quantum of diffusion, i.e. the minimally diffusive process. (c) When Thouless energy exceeds the energy level spacing, $E_T>\Delta$, the wave packet becomes extended and the usual diffusive behavior is recovered. Corresponding deformation potentials that carriers propagate on are shown in the background; grayscale for the static and yellowscale for the dynamic potentials.}
\label{thouless}
\end{figure}

In addition to our numerical results, quantum diffusive dynamics and its connection to the Planckian scattering rate can be understood phenomenologically by invoking the Thouless scaling theory~\cite{Edwards_1972,scaling}. Comparison of the Thouless energy with the emergent level spacing under the deformation potential characterizes the transport regimes in terms of Thouless metal-insulator transition phenomenology (described in Materials and Methods \ref{s:Pheno}). If the underlying lattice vibrations were frozen, the wave packet could have been localized due to coherent and strong interactions, Fig. \ref{thouless}(a). In reality, the deformation potential is time-dependent under which the wave packets can either be extended (Fig. \ref{thouless}(c)) or quantum diffusive (Fig. \ref{thouless}(b)). Although quantum diffusion can only occur at the equality of the characteristic energies, this regime can be robust rather than delicate because the deformation potential dynamics eventually cause delocalization regardless of how strong its magnitude is under reasonable ranges. Hence, the scaling theory provides a physical explanation for the observed numerical results.

\section*{Other strange metal puzzles}

In addition to giving rise to the $T$-linear resistivity at the Planckian threshold, the paradigm of quantum acoustics also sheds light on other perplexing aspects of strange metals. 

\begin{figure}[t]
\includegraphics[width=0.48\textwidth]{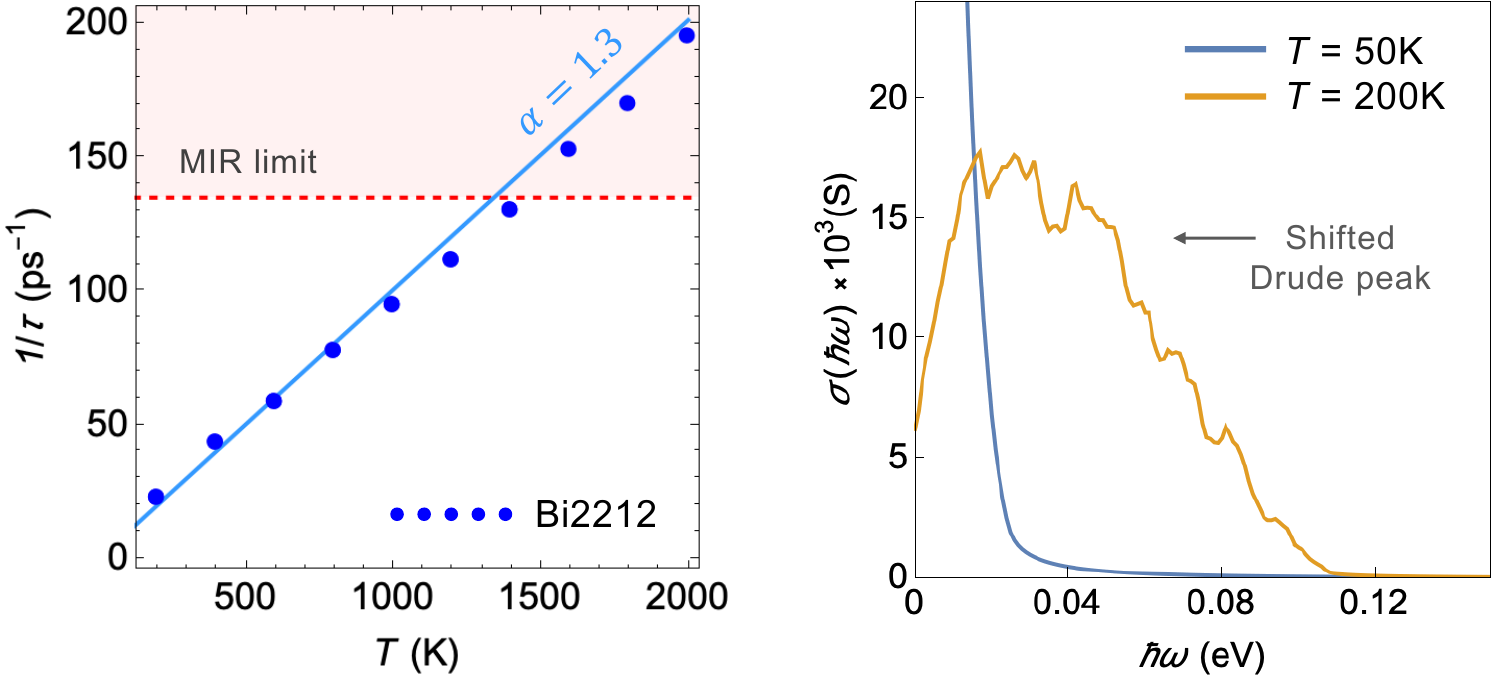}
\caption{Other strange metal mysteries. (a) The violation of the MIR limit for Bi2212 shows no sign of saturation. (b) The emergence of a displaced Drude Peak in the optical conductivity of YBCO at $T= 200\, \textrm{K}$. The increase of temperature drives the optical conductivity towards the non-Drude behavior: Instead of a peaking near zero-frequency, the optical conductivity exhibits a finite frequency maximum in the far infrared range, and the d.c. current is suppressed.
\label{Fig:DDP}
}
\end{figure}

First, at high temperatures, some highly-resistive, or simply ``good'' materials deviate from the expected $T$-linear behavior, which becomes progressively weaker as their resistivity approaches a constant value~\cite{RevModPhys.75.1085}. However, strange metals surpass the MIR bound with impunity at high temperatures~\cite{tahir-kheli_resistance_2013}. In Fig.~\ref{Fig:DDP}(a), we numerically demonstrate that the inverse scattering rate of Bi2212 (chosen because of its lower MIR limit compared to the other two) shows no saturation after passing by the MIR limit. While the linearity is preserved up to $2000\,\textrm{K}$ temperature, we observe a slight variation in the slope.
The standard narrative about the MIR limit relies on electrons acting like incoherent particles, which is essentially a semiclassical idea, not acknowledging the coherence of a charge carrier between one atom and the next. Our counter-argument borrows from the ``window glass" effect, wherein light traverses a landscape of scatterers roughly on a scale shorter than a wavelength, in Huygens-like multiple scattering regimes, producing unscathed forward scattering except for a lowered group velocity. Similarly to the window glass transparency, the electrons dispersed by the deformation potential bypass the MIR limit within the coherent state formalism of lattice vibrations. On the other hand, in the good-metal regime, we see that the deformation potential picture coincides with the standard explanation of this MIR saturation: the resistivity of a metal saturates since the mean free path cannot be smaller than the lattice constant, i.e. the smallest characteristic length scale in a material. Nevertheless, further analysis of the microscopic physics behind the phenomenon is required for a complete resolution of the mystery.

Another major mystery has been the displaced Drude peak effect (DDP). The emergence of a DDP has been attributed to an elusive dynamical, self-induced disorder that generates transient localization, hampering but not precluding charge carrier diffusion.\cite{infrared, infrared2}. The constantly  morphing deformation potential landscape generated by the conventional Fr\"ohlich Hamiltonian in the coherent state representation seems to us to be not only a natural but almost unavoidable \emph{identifiable} candidate. Unavoidable, in the sense that it is present, whether it is welcome or not. We welcome it, and suggest it is responsible for the DDP phenomena.  Fig.\ref{Fig:DDP}(b) shows an optical conductivity computed within the Kubo formalism in the presence of deformation potential, demonstrating a shift of the Drude peak towards a finite frequency maximum while the d.c. conductivity being suppressed with increasing temperature. Whereas the detailed study is presented in Ref.~\cite{PhysRevLett.132.186303}, the DDP is in a nutshell identified to emanate from (slowly) fluctuating degrees of lattice freedom, attributed to the same localization-delocalization interplay having a key role in the Planckian dynamics that we have found. A key property of the quantum-acoustical DDP is its signature temperature dependence that is comparable to the features reported in various families of strange metals, like in Ref.~\cite{infrared2}, but also distinguishes it from other types of DDPs, for instance, induced by impurities. 

Apart from MIR violation and DDP, we have strong hints about possible explanations of charge density wave and pseudogap regimes, which will be topics of future research. In particular, our preliminary calculations suggest that when electron motion strongly couples and synchronizes with low-energy lattice vibration modes, through Fr\"olich electron-phonon interaction treated nonperturbatively, a favorable environment for incommensurate charge density order could be created. Likewise, a similar resonance promoted by the deformation potential could result in temperature-dependent pseudogaps, i.e., a substantial suppression in the density of the low-energy excitations, which eventually melt away leaving the pseudogap phase regime. 
  
\section*{Discussion}

To summarize, near the critical doping and above the critical temperature in strange metals, charge carriers characterized by a slow, emergent velocity and an elevated effective mass, undergo intensified and ubiquitous interactions with lattice vibrations, leading to nonperturbative and phase preserving scattering in real space. Were the lattice vibration field static, this would lead to carrier localization~\cite{scaling}. However, the dynamic nature of the lattice vibrations counteracts the effects of their spatial strength and disorder, thereby inhibiting localization. This balance between static and dynamic properties of the lattice gives rise to a critical transport regime characterized by a minimum diffusion coefficient dictated by quantum limitations. Utilizing conventional transport theory, we demonstrate that this quantum diffusion is directly linked to a Planckian scattering rate, thereby providing both a prediction and an explanation for the observed Planckian resistivity in strange metals.

In addition to strange metals, many ``ordinary'' metals operate at the Planckian limit~\cite{strangemetal1,stabilitybound}. In fact, Peierls was the first to put forward the formula, which we now call the Planckian rate, by employing the perturbation theory of lattice-electron interaction for metals where the number of conduction electrons matches the chemical valency and their mass aligns with that of free electrons~\cite{Peierls1934}. He explicitly suggested replacing the perturbation method with a rigorous investigation of the time behavior of eigenfunctions (referred to as wavepackets in this context). After 90 years, we have realized it here with the emergence of quantum acoustics.  

The present quantum-acoustic route to linear and universal resistivity in strange metals has revealed an unexplored path unrelated to quantum criticality~\cite{sachdev1999quantum, Rev.Mod.Phy_94_035004_2022}, and not relying on (strong) electron-electron interaction~\cite{science_381_790_2023, sachdev2023quantum}, instead starting with the standard Fr{\"o}hlich Hamiltonian. In the broader context of strange metal features, there are models involving  self-generated randomness~\cite{infrared2} or a slowly moving bosonic field~\cite{scipost.phys_11_39_2021}. Even a recently published theory for strange metals~\cite{science_381_790_2023} evolves around a spatially random bosonic field. Here, we have demonstrated the deformation potential, i.e., lattice vibrations to be a strong candidate for this pivotal bosonic ingredient.

While neglecting the electron-electron correlations in this work is a strong assumption, there are reasons for electron-phonon coupling to dominate over electronic coupling in the strange metal regime above the critical temperature. In the famous strange metal phase diagram, electronic correlations are often associated with their proximity to the Mott (antiferromagnetic order) transition due to the local Coulomb interaction between the charge carriers~\cite{RevModPhys.94.035004}. With increased doping and temperature, electronic correlations (in the strange metal region) become less prominent, whereas phononic interactions become more appreciable~\cite{behnia}. On the other hand, closer to the Fermi liquid boundary, the contributions of electronic correlations are often incorporated in the Landau parameters, such as the effective mass. Planckian scattering seems to arise from a universal mechanism, ignoring the detailed nature and strength of the electron correlations in various materials~\cite{Legros2018}, supporting our universal quantum diffusion arguments. Nevertheless, our current results do not explain the continuation of the same $T$-linear slope down to very low temperatures, where the superconductivity is suppressed by the magnetic field~\cite{scalemagnetic,magnetic_field}. Below 50K, bare lattice vibrations seem insufficient to induce quantum diffusive behavior, leading to deviations from the Planckian slope at very low temperatures. Incorporating other mechanisms, such as back-action, magnetic fields, and electron-electron interactions, would be necessary to accurately predict low-temperature resistivity. Reconsidering the electron-electron correlations within our novel formalism might shed more light on the low-temperature regime. Protection of the resistivity slope against various interactions suggests to a universal speed limit.

Because the transport behavior of strange metals is ruled by the Planckian scattering rate, we focused here only on short time scale behavior in our system. At longer time scales, carrier coherence will be broken and the back-action of lattice vibrations on charge carriers as well as the thermalization process of charge carriers may have important consequences, especially in charge density wave and pseudogap regimes. In Appendix.~\ref{s:Backaction}, we reason that these back-action effects along with polaron instabilities can be justifiably discarded in the scope of this work. A further consideration of the physics of the back-action will be done in another paper.

In conclusion, foregoing the counting of phonons in favor of a quasiclassical description of the lattice reveals  important new physical insights and implications for strange metals. We have reinstated the role of electron-phonon interactions in strange metals, long thought to be unimportant. Any complete theory of carrier transport must consider phonon scattering above $T_c$, as it consistently manifests regardless of the relative significance of other scattering mechanisms. Our work affirms the existence of a bound in diffusivity of charge carriers and correctly predicts the universal Planckian resistivity in addition to shedding light on the possible origin of the violation of the MIR limit and the displaced Drupe peak. 




\section*{Materials and Methods}

\subsection*{Wave-on-Wave method}\label{s:Wow}

Inspired by quantum optics~\cite{Sudarshan1963,glauber1963,walls2007quantum}, the coherent state representation of lattice vibrations (phonons) has recently been developed to describe their interactions with charge carriers and applied to typical metals~\cite{Heller22}. Although the number state and coherent state pictures are formally equivalent, the approximations that they inspire are vastly different~\cite{PhysRevLett.10.277,Heller22}. In the perturbative limit, the coherent state description is equivalent to the conventional Fock state description. However, for nonperturbative   interactions, the coherent state picture reveals charge carrier coherence  beyond  single collision events, leading to transient localization~\cite{Heller22}.

The coherent state formalism describes the scattering events as quasielastic, coherence-preserving collisions with the moving hills and valleys of the quasiclassical field of lattice vibrations, i.e. the deformation potential. The hills and valleys act like a defect field, even in a perfect crystal, with the twist that the defects are constantly morphing from one shape to another. A lattice deformation is characterized by a displacement field $\mathbf{u}(\mathbf{x})$, the displacement of an atom at a position $\mathbf{x}$ from its equilibrium position. To represent electron-longitudinal acoustic phonon interaction under the coherent state description of lattice vibrations, we start with the quantum displacement field~\cite{Many-Particle}
\begin{eqnarray}\label{displament_field}
\begin{aligned}
    \hat{\mathbf{u}}(\mathbf{x},t)
    &=
    i\sum_{\mathbf{q}}\sqrt{\frac{\hbar}{2\rho_m \mathcal{V}\omega_{\mathbf{q}}}}
    \\
    &\qquad\qquad\quad\times
    \left( a_{\mathbf{q}}e^{-i\omega_{\mathbf{q}}t}
    +
    a_{-\mathbf{q}}^\dag e^{i\omega_{\mathbf{q}}t} \right)
    e^{i\mathbf{q}\cdot\mathbf{x}},
    \label{eq:uquantum}
\end{aligned}
\end{eqnarray}
where $\mathbf{q}$ is wave vector of a normal mode, $\omega_{\mathbf{q}}$ is phonon frequency, $\rho_m$ is mass density, $\mathcal{V}$ is volume (area in 2D), and $t$ is time. We omit writing the polarization vector and phonon branch indices in the subscripts since we only deal with the longitudinal acoustic modes of the lattice vibrations, the main scattering mechanism for the charge carriers.

We employ the deformation potential approach as the first-order correction in the expansion of electronic band energy as a function of strain fields due to atomic displacements~\cite{BS2}. Then the quantum field of the deformation potential is written as
\begin{eqnarray}
\begin{aligned}
    \hat{V}_D(\mathbf{r},t)
    &=
    E_d\nabla\cdot\hat{\mathbf{u}}(\mathbf{r},t)
    \\
    &=
    -\sum_{\mathbf{q}}
    g_{\mathbf{q}}
    (a_{\mathbf{q}}e^{-i\omega_{\mathbf{q}}t}
    +
    a_{-\mathbf{q}}^\dag e^{i\omega_{\mathbf{q}}t})
    e^{i\mathbf{q}\cdot\mathbf{r}},
    \label{VDquantumfield}
\end{aligned}
\end{eqnarray}
where the parameter $g_{\mathbf{q}}=E_d
\sqrt{\hbar/(2\rho_m \mathcal{V}\omega_{\mathbf{q}})}
\abs{\mathbf{q}}$ represents electron-phonon coupling strength.

Coherent state description allows us to treat phonons as waves and construct a quasiclassical lattice vibration field yielding a deformation potential that acts as a real internal field on charge carriers. Each normal mode of lattice vibration with a wave vector $\mathbf{q}$ is associated with a coherent state $\vert \alpha_{\mathbf{q}} \rangle$. At thermal equilibrium, each mode is considered to be equilibrated with a heat bath at temperature $T$, giving thermal ensembles of coherent states~\cite{BS2}. Unlike a Fock state, a coherent state contains an indefinite number of quanta. However, in equilibrium, a thermal coherent state has an average occupation, $\langle n_{\mathbf{q}}\rangle_{\textrm{th}}=1/[\exp\left(\hbar\omega_{\mathbf{q}}/k_{\textrm{B}} T\right) - 1]$, given by Bose-Einstein statistics. Then, a thermal coherent state is defined as
\begin{equation}
    \alpha_\mathbf{q}=\sqrt{\langle n_{\mathbf{q}}\rangle_{\textrm{th}}}\exp(i\varphi_\mathbf{q}),
\end{equation}
where $\sqrt{\langle n_{\mathbf{q}}\rangle_{\textrm{th}}}=\vert\alpha_\mathbf{q}\vert$ is the thermal amplitude and $\varphi_{\mathbf{q}}=\arg(\alpha_{\mathbf{q}})$ is the random phase of the coherent state determining the initial conditions. Using the independence of normal modes, entire lattice vibrations can be described as the product state of the coherent states $\ket{\alpha_{\mathbf{q}}}$ of the normal modes $\mathbf{q}$'s, a multimode coherent state~\cite{hellerkim}
\begin{eqnarray}
\begin{aligned}
\ket{\bm{\alpha}}=\prod_{\mathbf{q}}\ket{\alpha_{\mathbf{q}}}. 
\end{aligned}
\end{eqnarray}

Now to construct the quasiclassical field of lattice vibrations, we take the expectation value of the quantum field of the deformation potential with respect to the multimode coherent state $\ket{\bm{\alpha}}$:

\begin{eqnarray}\label{quasiclassical_DP}
\begin{aligned}
    V_D(\mathbf{r},t)
    &=\bra{\bm{\alpha}}\hat{V}_D(\mathbf{r},t)\ket{\bm{\alpha}}
    \\
    &=
    \sum_{\mathbf{q}}^{q_D}
    g_{\mathbf{q}}
    (\alpha_{\mathbf{q}}e^{-i\omega_{\mathbf{q}}t}
    +
    \alpha_{-\mathbf{q}}^* e^{i\omega_{\mathbf{q}}t})
    e^{i\mathbf{q}\cdot\mathbf{r}},
\end{aligned}
\end{eqnarray}
which gives rise to the quasiclassical field of deformation potential. We omit writing the negative sign of the potential because it is statistically the same upside down. We use the Debye model, introducing the linear dispersion $\omega_{\mathbf{q}}=v_s|\mathbf{q}|$ where $v_s$ is sound speed. Our formalism follows step by step from a Fr\"olich Hamiltonian without stopping for any new assumptions, except (1) we take the coherent state, the classical-like limit for the field, and (2) we neglect back-action of the carrier on the lattice, which is not important in most of the temperature ranges investigated here. More information can be found in Ref.~\cite{Heller22}. 

Even though the deformation potential is composed of many thermally occupied normal modes and does not possess a single characteristic length or time scale, its dynamics can mainly be determined by its largest frequency components. Below the Debye temperature, the deformation potential timescale is $\tau_D \sim 2\pi/\omega(T)$ where $\omega(T)$ is the temperature-dependent Bose frequency~\cite{Heller22}. Above the Debye temperature, $\omega(T)$ saturates to $\omega_D$ which is the Debye frequency. For typical metals, lattice vibrations move much slower compared to charge carriers, and at very short time scales ($\tau<\tau_D$) the deformation potential appears to be frozen. When the electron-phonon coupling is strong, the disordered landscape generated by frozen deformation potential could localize the carriers at short-time scales due to their coherent interactions with lattice vibrations even at high temperatures~\cite{Heller22}. However, the time-dependence of the deformation potential sets in at longer time scales ($\tau>\tau_D$) giving rise to transient localization~\cite{PhysRevB.83.081202,PhysRevB.89.235201} and eventual delocalization of carriers. See Ref.~\cite{Heller22} for a more detailed discussion and application of this approach.

Comparison of the energy of the carriers $E_F$ (usually Fermi energy) with the root mean square of the deformation potential $\langle\sqrt{V_D^2}\rangle$ (averaged over many realizations with different initial phases of coherent states) determines whether the scattering can be treated perturbatively or not, see Eq. (11a) below. Also, quantum coherence of charge carriers becomes important in scattering when the wavelength of the carriers $\lambda_F$ (Fermi wavelength) is not much less than the twice of the lattice constant $a$, see Eq. (11b). It should be emphasized that the term 'coherence' is employed here to describe the spatial phase coherence of the carrier wavefunction, and should not be conflated with the 'coherent versus incoherent metals' nomenclature, which pertains to the breakdown of the quasi-particle paradigm~\cite{Hartnoll2015}. In summary, we have
\begin{subequations}
\begin{align}
\bar{E} =\frac{E_F}{\langle\sqrt{V_D^2}\rangle} &
\begin{cases}
	\bar{E}\gg 1  & \rightarrow \;  \text{Perturbative} \\
	\bar{E}\sim 1 \; \text{or} \; \bar{E}\ll 1  & \rightarrow \;  \text{Nonperturbative},
\end{cases}
\\
\bar{\lambda} =\frac{\lambda_F}{2a} &
\begin{cases}
	\bar{\lambda}\ll 1  & \rightarrow \;  \text{Incoherent} \\
	\bar{\lambda}\sim 1 \; \text{or} \; \bar{\lambda}>1  & \rightarrow \;  \text{Coherent}.
\end{cases}
\end{align}
\end{subequations}

The motion of charge carriers in a material can be modeled by quantum wave packet propagation methods. In particular, under a time-dependent lattice potential, it is more convenient to use time-dependent wave packet propagation for the carrier as opposed to the carrier eigenstates. We focus on the quantum dynamics of a carrier under the following time-dependent Hamiltonian
\begin{align}
    H(\mathbf{k},\mathbf{r},t)
    &=
    E_0(\mathbf{k})
    +
    V_D(\mathbf{r},t),
    \label{DPHam}
\end{align}
where $\mathbf{k}$ is the charge carrier wave vector, $E_0(\mathbf{k})=\hbar^2\mathbf{k}^2/2m^*$ is the carrier energy with the effective mass $m^*$, and $V_D(\mathbf{r},t)$ is the deformation potential. The effect of an external magnetic field defined by a vector potential $\mathbf{A}$ by taking into account the level of minimum coupling is
\begin{equation*}
    E_0 = \frac{1}{2m^*} \left( \hbar \mathbf{k} + q\mathbf{A} \right)^2,
\end{equation*}
where $q$ is the charge of a carrier.

In our simulations, we employ the (third-order) split operator method for the propagation of a wavepacket~\cite{Tannor,Heller:way}, exploiting the highly optimized fast Fourier transformation algorithms. However, in the presence of a magnetic field, the method seems to lose its edge, since the minimum-coupling kinetic operator is no longer diagonal in the wave vector space. Nevertheless, when the magnetic field is homogeneous and parallel to the studied system, the standard split-operator method~\cite{Tannor,Heller:way} can be extended \emph{exactly}. By incorporating the exact factorization of the kinetic energy of Aichinger et al.~\cite{aichinger_comput.phys.commun_171_197_2005}, our implementation has paved the way for an (even very strong) homogeneous external magnetic field to be included in split-operator simulations with no additional approximations, and without compromising the original proficiency. Furthermore, our method gains an additional boost in its computational efficiency by choosing the linear gauge for the vector potential, but our method can be also designed~\cite{janecek_phys.rev.B_77_245115_2008} in a gauge-invariant manner regardless of the discretization, removing gauge-origin issues that often plague computations with a magnetic field.

For charge carriers, we consider wave packets having an average emerged momentum $\hbar k$ to the $\hat{x}$ direction
\begin{eqnarray*}\label{eq:initial_wavepacket}
\begin{aligned}
    \psi(0)=\frac{1}{2\pi\sigma_x\sigma_y}\exp( -\frac{(x-x_0)^2}{2\sigma_x^2}
    -\frac{(y-y_0)^2}{2\sigma_y^2}+ikx),
\end{aligned}
\end{eqnarray*}
on the deformation potential $V_D(\mathbf{r},t)$, with or without an external magnetic field.
The initial wave function $\ket{\psi(0)}$ is propagated with the propagator $U(t)$ to obtain the state $\ket{\psi(t)}=U(t)\ket{\psi(0)}$ at time $t$. Then, any observable is calculated as follows
\begin{align}
\langle\hat{\mathbf{O}}\rangle(t)=\mel{\psi(t)}{\hat{\mathbf{O}}}{\psi(t)}.
\end{align}
We calculate the change in wave packet position by $\langle\hat{\mathbf{x}}\rangle(t)=\mel{\psi(t)}{\hat{\mathbf{x}}}{\psi(t)}$ and the decay of carrier velocities by $\langle\hat{\mathbf{v}}\rangle(t)=\mel{\psi(t)}{\hat{\mathbf{v}}}{\psi(t)}$, where $\mathbf{x}$ is position and $\mathbf{v}$ velocity operators in the transport direction respectively.

We thus wind up solving two dynamically coupled wave equations (i.e. Schrödinger equation for charge carriers, and wave equation for the lattice) simultaneously in real space, which we call the WoW method. We determine the diffusion coefficients by extracting the slope of the mean square distance (function fitting and manual extraction give almost the same results since the function is linear). At the initial launch of the wavepacket, the spread is non-linear. Then it quickly saturates to a linear slope, where we extract the diffusion coefficients. To prevent the influence of simulation box size effects, as well as lattice back action, we terminate our simulations as early as possible after the initial non-linear spread of the wave packet and as soon as it saturates to a linear slope. Based on our lattice back action and energy dissipation considerations, the quantum coherence of charge carriers is eventually broken after around 0.3 ps, which agrees with the temporal endpoint for our simulations. In other words, simulating longer time scales won't be meaningful anyway due to the incoherent transport regime developing. The transport of carriers is determined by the coherent, short-timescale behavior as evident from the Planckian scattering rate. To eliminate the influence of the initial conditions on the obtained diffusion rates, we employ disorder averaging across 10 distinct realizations of the deformation potential.

\subsection*{Strange metal parameters}\label{s:Parameters}

In Table \ref{table}, we present the empirical data for the material parameters used in this work for the considered strange metals at critical doping. To check the robustness of the quantum diffusion, we consider the following material parameter ranges for generic strange metals at the critical doping: The longitudinal sound velocity ($v_s$) can be as low as $2.7\times 10^3$ m/s~\cite{PhysRevB.75.144511,PhysRevB.106.235134} and can reach up to $5\times 10^3$ m/s~\cite{PhysRevB.52.R13134,SMconstants,PhysRevB.100.241114}. $v_s$ not only characterizes the movement of the lattice vibrations but also indirectly determines the emergent velocity of the carrier. In-plane lattice parameters vary between $3.8$ \AA ~\cite{PhysRevLett.89.107001} and $5.4$ \AA ~\cite{BABAEIPOUR2005130}. The hole effective mass values range between 3 to 10 times the bare electron mass~\cite{PhysRevB.72.060511,Legros2018}. In general, the typical range for the deformation potential constant, $E_d$, is between $1$ eV to $40$ eV. We choose $E_d=10$ eV as a reasonable value for all cuprates. The variations of the interplane spacing of adjacent CuO$_2$ planes and the mass density can also be accounted for within the range of $E_d$ values. Finally, the material-dependent factor $\alpha$ in the Planckian rate predominantly varies between 0.5 and 2.0 for a wide range of strange metals~\cite{strangemetal1,Legros2018,Tlinearresistivity,magnetic_field}.

\begin{table}[t]\label{table}
\caption{The material parameters for three different strange metals.}
\def\arraystretch{1.5}
\setlength{\tabcolsep}{0.5em}
\begin{tabular}{llllll}
Material   & $m^* (m_0)$ & $v_s$ (m/s)  & $a$ (\AA)  \\ \hline
LSCO                   & 9.8         & 4800                                           & 3.8            \\
Bi2212                      & 8.4         & 2800                                           & 5.4              \\
YBCO                    & 3.0         & 5850                                           & 3.8              \\ \hline
\end{tabular}
\label{table}
\end{table}

\subsection*{Phenomenological model}\label{s:Pheno}

In the main text, we have shown the numerical evidence of quantum diffusion in strange metals due to the strong deformation potential scattering of carriers. The diffusion of charge carriers in the interplay of localization and delocalization can be understood by invoking the phenomenological model of Thouless scaling theory which has been developed to explain the metal-insulator transition in disordered systems~\cite{Edwards_1972,scaling}. Within this phenomenological model, the metal-insulator transition is understood by comparing the typical energy levels spacing due to the underlying disordered potential with the required energy of wave packet to diffuse. The latter one is called Thouless energy, the energy scale determining metal-insulator transition in disordered metals, as well as in strongly correlated systems~\cite{PhysRevB.73.184515}.

In our simulations, we observe that wave packets rapidly lose their initial velocity and form quasibound states for a short time after their initial launch. Their energy spectrum resembles to the ones of localized states due to strong deformation potential. The mean level spacing of states is defined as
\begin{equation}
    \Delta=\frac{1}{g(E)L^d},
\end{equation}
where $g(E)$ is the density of states and $d$ is the dimension of the system with an associated length $L$. Nearly bound states of carriers live in so-called Thouless blocks, whose size is determined by the approximate localization length of the wave packet. For a 2D system, $g(E)=m^{*}/(\pi\hbar^2)$ and the typical energy level spacing within the Thouless block is given by
\begin{equation}
    \Delta\sim\frac{\hbar^2}{m^{*}L^2},
\end{equation}
Since each Thouless block is basically open and the underlying deformation potential is dynamic, wave packets diffuse through the boundaries of the blocks within the timescale $\tau_T=L^2/D$. The corresponding energy due to diffusion, i.e. Thouless energy,
\begin{equation}
    E_T=\frac{\hbar D}{L^2}.
\end{equation}

Metal-insulator transition is determined by the comparison of these energy (or corresponding time) scales. The wave packet is localized when it does not have enough energy to overcome the energy level spacing ($\Delta>E_T$) of the nearest block and is extended when the conditions are reversed, see Fig. \ref{thouless}. However, incessant competition between  localization and delocalization keeps the wave packet on the edge of the critical coupling and drives it to the quantum bound of diffusion. At this critical point where $\Delta=E_T$, $L$'s cancel out from both sides and the diffusion constant becomes equal to $D_Q=\hbar/m^{*}$. This is the minimum diffusion a quantum system can have, representing the lower bound. Below this value, the system is localized. 

The reason we observe the quantum diffusion limit is that the deformation potential coupling is so strong that the underlying dynamics just barely delocalize the wave packet, driving the system into the quantum diffusion conditions. Note that, the dimensionality of the system also plays a role here. It is easier to get localized in a 1D system and harder in a 3D one. It turns out that the 2D nature of the transport makes it favorable for the system to saturate to the quantum diffusivity bound.

\subsection*{Lattice back-action}\label{s:Backaction}

In this work, an electron feels a rolling potential landscape stemming from the lattice distortions but does not influence back onto the deformation potential. This kind of asymmetrical interaction can lead to, for instance, unphysical heating of the electron in the long run, as the potential ''pumps'' energy into the electron motion, but the electron cannot ''dump'' energy back to the lattice (this does not happen here, because we stop the simulations before the quantum phase coherence breaks down). For completeness, we want to point out that the time-reversal process does exist: the lattice ``drains" the electron; whereas the electron does not gain energy back from the lattice. This fact should not be confused that the deformation interaction is thus bilateral. In addition, this approach omits the effect of the lattice distortion caused by the electron on its on motion, which is a key feature, e.g., in the polaron formation. The former issue is numerically established to be insignificant in our simulation time frame. For the latter, we here argue that the presented results are valid, and the back-action plays a minor role at high enough temperatures, which we estimate below.

From the mathematical standpoint, the deformation potential is the coherent state description of lattice vibrations taken to the quasiclassical field limit, which originates from the conventional Fröhlich Hamiltonian~\cite{Frochlich1, Frochlich2} encompassing the lowest-order (linear) lattice-electron coupling~\cite{Mahan}:
\begin{equation*} 
\mathcal{H}_{\textrm{F}} = \frac{\mathbf{p}^2}{2 m^*} + \sum_{\mathbf{q}} \hbar \omega_{\mathbf{q}} a_{\mathbf{q}}^{\dagger} a_{\mathbf{q}} + \sum_{\mathbf{q}} \Big( V_{\mathbf{q}} a_{\mathbf{q}}e^{i \mathbf{q} \cdot \mathbf{r}} + \textrm{h.c.} \Big),
\end{equation*}
where $\mathbf{r}$ is the position coordinate operator of the electron with effective (band) mass $m^*$, $\mathbf{p}$ is its canonically conjugate momentum operator; $a_{\mathbf{q}}$ $(a_\mathbf{q}^{\dagger})$ is the creation (annihilation) operator for longitudinal optical phonons of wave vector $\mathbf{q}$ and energy $\hbar \omega_{\mathbf{q}}$. The electron-phonon interaction is defined by its Fourier components $V_{\mathbf{q}}$, which are sometimes expressed through a dimensionless coupling parameter determined by the dielectric constants of a polar crystal, or as here, they are directly linked to our couplings $g_{\mathbf{q}}$.

Employing the coherent state formalism detailed in Ref.~\cite{Heller22}, our Hamiltonian beget a deformation potential composed of thermalized lattice modes $\vert \alpha_{\mathbf{q}}(t=0) \vert = \sqrt{n_{\mathbf{q}}}$, causing the electron to undergo quasielastic, coherence-preserving scattering events. In return, an electron gives feedback on the amplitudes $\alpha_{\mathbf{q}}(t)$ while roaming in the malleable landscape of rising and falling electrostatic hills and valleys. However, the back-action contribution on the deformation potential is almost of no consequence if all the electron-lattice couplings related to the allowed lattice modes are small compared to the corresponding mode energies, i.e.,
\begin{equation}\label{eq:BA1}
    \frac{g_{\mathbf{q}}}{h\omega_{\mathbf{q}} \sqrt{n_{\mathbf{q}}}} \ll 1 \quad \textrm{ for all modes} \quad \vert \mathbf{q} \vert \le q_D.
\end{equation}
By considering the worst-case scenario, the criteria in Eq.~\ref{eq:BA1} corresponds to temperatures of order of
\begin{equation}\label{eq:BA2}
\frac{T_D}{T} \ge \ln \Big[ \Big(\frac{g_{\textrm{max}}}{\hbar \omega_{\textrm{min}}} \Big)^2 + 1 \Big].
\end{equation}
In the equation above, we have defined the maximum coupling $g_{\textrm{max}}$ and the minimum vibration angular frequency $\omega_{\textrm{min}}$ supported by the lattice as
\begin{equation*}
    g_{\textrm{max}} = E_d \sqrt{\frac{\hbar q_D}{2 \rho_m \mathcal{V} v_s}}
    \quad \textrm{and} \quad
    \omega_{\textrm{min}} = \hbar v_s \frac{\pi}{L},
\end{equation*}
where we have assumed the conventional linear dispersion for acoustic phonons, and $L$ is the characteristic length scale of the system in the sense of $\mathcal{V} = L^d$ ($d$ being the dimension of the system).
 
For the considered cuprates, the estimation in Eq.~\ref{eq:BA2} yields a back-action temperature of $T_{\textrm{BA}} \sim 80\, \textrm{K}$. This analysis is in line with the presented results: small deviations from the Planckian rate is seen in Fig.~\ref{planckian} around the temperatures below the back-action estimation $T_{\textrm{BA}}$ given by Eq.~\ref{eq:BA2}. Above this temperature, the carrier's effect on the deformation potential dies out in an exponential rate, which is captured by our preliminary simulations including the full lattice-electron dynamics showing that the results presented here are well-founded, and further validating our approximation of excluding the back-action.  

\subsection*{Kubo calculations}\label{s:DDP}

The linear response theory pioneered by Kubo is directly applicable to determining the optical conductivity of charged particles. In this formalism~\cite{kubo_formalism}, the (complex) conductivity tensor $\sigma_{\mu \nu}(\omega)$ relates the current density $j_{\mu}(t)$  in the $\mu$-direction to the applied electric field $\Re[E_{\nu} \exp(i\omega t)]$ in the form of
\begin{equation*}
    j_{\mu}(t) = \Re[ \sigma_{\mu \nu}(\omega)  E_{\nu} \exp(i\omega t)].
\end{equation*}
If the system is in thermal equilibrium with a heat reservoir at temperature $T$, the conductivity tensor is given by the Kubo formula,
\begin{equation*}\label{eq:Kubo_origin}
    \sigma_{\mu \nu}(\omega) = \lim_{\eta \rightarrow 0+}\frac{\mathcal{V}}{\hbar \omega} \int_{0}^{\infty} \textrm{d}t\, \Big\langle [\hat{j}_{\mu}(t), \hat{j}_{\mu}(0)] \Big\rangle e^{-i(\omega  + i \eta)t},
\end{equation*}
where $\mathcal{V}$ is the volume of the system, and $\langle \cdots \rangle$ refers to the averaging over equilibrium thermal ensemble. Hereon, we are only interested in one diagonal component of the conductivity tensor, namely $\sigma := \sigma_{xx}$. 

When we first freeze the deformation potential, we determine a set of electronic eigenstates $\vert n \rangle$, and then compute the corresponding optical conductivity as
\begin{equation*}
\begin{split}
\sigma^{(F)}(\omega) = -\lim_{\eta\to0^+}\frac{e^2 \hbar}{\mathcal{V}{m^*}^2} &\sum_{n,m} \frac{f(\varepsilon_n) - f(\varepsilon_m)}{\varepsilon_n - \varepsilon_m}\\
&\times\frac{|\bra{n}\hat{p}_x\ket{m}|^2}{(\hbar\omega+\varepsilon_n - \varepsilon_m)^2+\eta^2}\eta,
\end{split}    
\end{equation*}
where $\hat{p}_x$ is the momentum operator and $m^*$ the effective mass. In the dynamic scenario of a morphing deformation potential, we utilize the solved eigenstates from the frozen approximation by treating them as initial states and establish the optical conductivity according to the Kubo formula as
\begin{equation*}\label{eq:dynamic_formula}
    \begin{split}
    \sigma^{(D)}(\omega) = -&\lim_{\eta\to0^+}\frac{e^2}{\mathcal{V}{m^*}^2} \sum_{n,m} \frac{f(\varepsilon_n) - f(\varepsilon_m)}{\varepsilon_n - \varepsilon_m}\\
    &\times\int_0^\infty\mathrm{d}t\bra{n}\hat{p}_x(t)\ket{m}\bra{m}\hat{p}_x\ket{n}e^{i(\omega+i\eta)t},
    \end{split}
\end{equation*}
where the momentum operator $\hat{p}_x(t)$ is expressed in the Heisenberg picture. This two-step strategy enables us to analyze the effect of the dynamics associated with the potential landscape by assessing the optical conductivity with the frozen approximation $\sigma^{(F)}$ and under the full time-evolution $\sigma^{(D)}$.

\begin{acknowledgments}
We thank J. H. Miller Jr., F. Giustino, S. Das Sarma, B. I. Halperin, A. P. MacKenzie, S. Datta, D. Kim, and A. M. Graf for useful discussions. We thank X.~Ouyang and S.~Yuan for their improvements in the Wave-on-Wave code. We thank the National Science Foundation for supporting this research, through the NSF the Center for Integrated Quantum Materials (CIQM) Grant No. DMR-1231319. J.K.-R. thanks the Emil Aaltonen Foundation and the Oskar Huttunen Foundation for financial support.
\end{acknowledgments}

\vspace{20pt}

\textbf{Author Contributions:} All authors contributed to the development of the idea and the writing of the manuscript. A.A. conceived the idea for the phenomenological quantum diffusion model, extracted the empirical data, performed the numerical calculations, and wrote the original draft. J.K.-R. contributed to the back-action argumentations and performed the displaced Drude peak and Kubo calculations. E.J.H. supervised the work.

\bibliography{refs}






\end{document}